\documentclass{camera}
\usepackage{graphicx}  

\begin{document}

%
\title{The Sooner: a Large Robotic Telescope}

%
\author{G. Chincarini$^{1,2}$, M. Zannoni$^{1}$, S. Covino$^{2}$, E. Molinari$^{2,3}$, S. Benetti$^{4}$,
F. Vitali$^{5}$, C. Bonoli$^{4}$, F. Bortoletto$^{4}$, E. Cascone$^{6}$, R. Cosentino$^{3}$, 
F. D'Alessio$^{5}$, P. D'Avanzo$^{1,2}$, V. De Caprio$^{2}$, M. Della Valle$^{6}$, A. Fernandez-Soto$^{7}$, 
D. Fugazza$^{2}$, E. Giro$^{4}$, A. Gomboc$^{8}$, C. Guidorzi$^{9}$, D. Magrin$^{4}$, G. Malaspina$^{2}$, L. Mankiewicz$^{10}$, 
R. Margutti$^{1}$, R. Mazzoleni$^{2}$, L. Nicastro$^{11}$, A. Riva$^{12}$, M. Riva$^{13}$, 
R. Salvaterra$^{1}$, P. Span\`o$^{2}$, M. Sperandio$^{2}$, M. Stefanon$^{7}$, G. Tosti$^{14}$ \and V. Testa$^{5}$}

%
\small{\organization{
$^{1}$\,Dipartimento di Fisica G. Occhialini, Universit\`a degli Studi di Milano - Bicocca, Piazza della Scienza 3, 20126 Milano, Italy \\
$^{2}$\,Istituto Nazionale di Astrofisica, Osservatorio Astronomico di Brera, via Bianchi 46, 23807 Merate, Italy\\
$^{3}$\,Istituto Nazionale di Astrofisica, Telescopio Nazionale Galileo, Rambla Jos\'e Ana Fern\'andez P\'erez, 7, 38712 Bre\~na Baja, Spain\\
$^{4}$\,Istituto Nazionale di Astrofisica, Osservatorio Astronomico di Padova, Vicolo dell'Osservatorio, 5, 35122 Padova, Italy\\
$^{5}$\,Istituto Nazionale di Astrofisica, Osservatorio Astronomico di Roma, Via Frascati, 33, Monte Porzio Catone 00040 Roma, Italy\\
$^{6}$\,Istituto Nazionale di Astrofisica, Osservatorio Astronomico di Capodimonte, Salita Moiariello, 16, 80131 Napoli, Italy\\
$^{7}$\,Instituto de Fisica de Cantabria (CSIC-UC), Edificio Juan Jorda, Av. de los Castros s/n, 39005 Santander, Spain\\
$^{8}$\,Fakulteta za matematiko in fiziko, Univerza v Ljubljani, Jadranska 19, 1000 Ljubljana, Slovenia\\
$^{9}$\,Dipartimento di Fisica, Universit\`a di Ferrara via Saragat 1, 44122 Ferrara, Italy
$^{10}$\,Center for Theoretical Physics of Polish Academy of Science, Al. Lotnikow 32/46, 02-668 Warsaw, Poland\\
$^{11}$\,Istituto Nazionale di Astrofisica, IASF Bologna, via Gobetti 101, 40129 Bologna, Italy\\
$^{12}$\,Istituto Nazionale di Astrofisica, Osservatorio Astronomico di Torino, Via Osservatorio, 20, 10025 Pino Torinese, Italy\\
$^{13}$\,Dipartimento Ingegneria Aerospaziale, Politecnico Milano, Via La Masa, 34, 20156 Milano, Italy\\
$^{14}$\,Dipartimento di Fisica, Facolt\`a di Scienze MM. FF. NN., Universit\`a degli Studi di Perugia, via A. Pascoli, 06123 Perugia, Italy\\
}}

\maketitle


%
The approach of Observational Astronomy is mainly aimed at the construction of larger aperture 
telescopes, more sensitive detectors and broader wavelength coverage. Certainly fruitful, this 
approach turns out to be not completely fulfilling the needs when phenomena related to the formation 
of black holes (BH), neutron stars (NS) and relativistic stars in general are concerned. Indeed they 
manifest themselves through highly variable emission of electromagnetic energy and quite often via 
sudden bursts of electromagnetic energy possibly accompanied (or preceded) by the emission of gravitational 
waves and neutrinos. These are expected to occur in the collapse of massive stars into GRBs or SNe to 
produce a BH or a NS, and in the merging of such relativistic objects (NS+NS and BH+NS).\\
Radio observations and later X ray observations showed that we are living in a violent and very dynamic 
highly variable Universe where the energy involved in explosive phenomena may be as large as $10^{52}$ - $10^{54}$ erg 
(isotropic energy). Since then the field developed theoretically thanks to the contribution of many 
gifted theoreticians and observationally thanks to a huge development in the area of instrumentation and detectors. 
Recently, mainly through the Vela, Beppo-­SAX and Swift satellites, we reached a reasonable knowledge of the most 
violent events in the Universe and of some of the processes we believe are leading to the formation of 
black holes (BH). Massive BHs are believed to exist in AGN and in the nuclear region of the galaxies in general.\\

We plan to open a new window of opportunity to study the variegated physics of very fast astronomical transients, 
particularly the one related to extreme compact objects. The innovative approach is based on three cornerstones: 
1) the design (the conceptual design has been already completed) of a 3m robotic telescope and related focal plane 
instrumentation characterized by the unique features: ``No telescope points faster''; 2) simultaneous 
multi-­wavelengths observations (photometry, spectroscopy \& polarimetry); 3) high time resolution observations. 
The conceptual design of the telescope and related instrumentation is optimized to address the following topics: 
High frequency a-­periodic variability, Polarization, High z GRBs, Short GRBs, GRB--Supernovae association, 
Multi-­wavelengths simultaneous photometry and rapid low dispersion spectroscopy. This experiment will turn 
the ``exception'' (like the optical observations of GRB\,080319B) to ``routine''.
\begin{figure}
\includegraphics[width=\hsize]{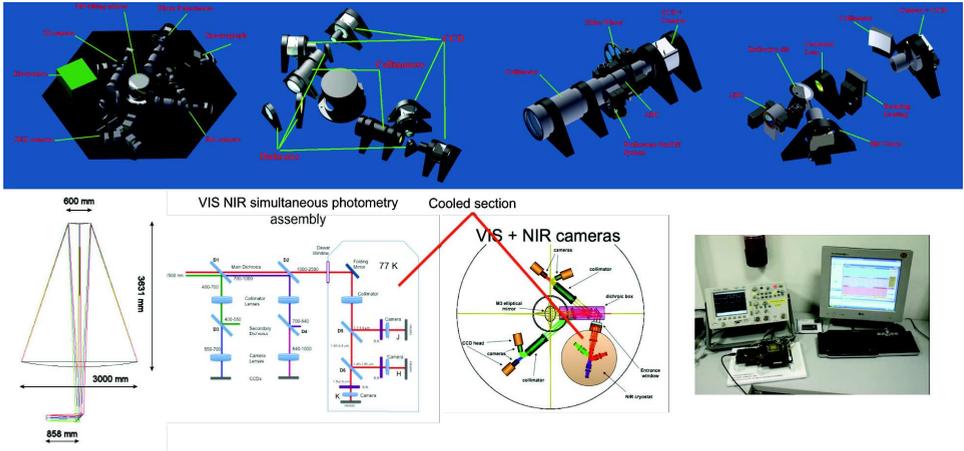}
\caption{Top: Focal plane instrumentation with all the instruments developed perpendicular to the optical 
axis of the telescope. Details on the single instruments and mounting of the optics are given in the exploded 
figures on the top right part of the figure. Each instrument has its own detector at the focal plane and all of 
them cover an optimum range of wavelengths in order to have the best coverage possible. The spectrograph in 
particular covers the wavelength range 370 -- 920 nm to identify easily and quickly the high z objects. The 
polarimeter will use either o double Wollaston prism or a system as the one developed for the Liverpool 
telescope by \cite{Steele06}. The telescope, bottom part of the figure, has an aperture of 300 cm and 
will carry all the instrumentation at the Cassegrain focus. The multi-­wavelength photometry will be carried out 
simultaneously with four CCD working at optical wavelengths (g, r, I, z) and three in the near infrared (NIR, J, H, K). 
The field of view of the cameras is of 10 x 10 arcmin$^{2}$ while with the spectrograph we plan to over a field of 
2 x 8 arcmin$^{2}$. For further details see \cite{Vitali10}.}
\label{fig01} 
\end{figure}
In GRB\,080319B (the naked eye GRB\,\cite{Racusin08}) we observed rapid variability both at high energy and in 
the optical. The optical light curve variability on a scale of 5 to 10 s follows the hard X ray variability 
with a delay of about 6 seconds. Differences between the optical and the high energy exist (here however it 
is essential to plan for higher time resolution at optical wavelengths) at higher frequency (variability). 
Invoking an inverse Compton for the high-­energy emission the differences in variability and the delay call 
for a revised model\,\cite{Guidorzi10}. In other words since the inverse Compton on the synchrotron photons 
occurs with no observable delay, the high energy signal should follow the optical prompt emission light curve 
in all the details and without any delay unless the sources are in different locations. The high speed 
and multi-­wavelengths photometry at optical wavelengths during the early phase of the prompt emission will 
lead to a major advance of our understanding of gamma-­ray burst, and help answer one of the fundamental 
unanswered questions as to how the radiation is produced in these explosions.\\

We need furthermore to understand about the magnetic field. Polarimetry during the prompt emission would be 
exceedingly important to determine the geometry and origin of magnetic fields in GRB shocks.\\

One of the long-­standing issues with our current understanding of long GRBs is that the supernovae associated 
with these bursts are type Ic supernovae (suggesting that long GRBs lack both hydrogen and helium atmospheres). 
The lack of a hydrogen atmosphere was expected. Hydrogen atmospheres are too extended for the jet to 
propagate through the star during the disk accretion timescale, leading to mildly relativistic jets. But it is 
believed that helium atmospheres are compact and most of the progenitors of long GRBs predict the outburst to 
be associated with both type Ib and type Ic supernovae. Either the progenitor for long GRBs requires a more exotic 
model than many of the current proposals, or the nature of the explosion has hidden the helium, making what would 
normally be a type Ib supernova appear as a type Ic supernova.\\
One clue to this long GRB progenitor problem lies in understanding shock break out. This could have been observed 
in GRB\,060218 and in 2008D/XRF\,080109\,\cite{Campana06,Mazzali08,Soderberg08}. On the other hand this 
interpretation is being debated since a similar phenomenon could be caused by a shockwave interacting with gas 
shells ejected by luminous blue variable outbursts. The complexity of the problem requires full 
radiation-­hydrodynamics calculations as those carried out by C. Fryer at Los Alamos\,\cite{Fryer07}; however 
for these calculations it is critical to have an observational counterpart to constrain the timing of such 
phenomena.\\
One of the major goals justifying the search of high z galaxies is, in addition to the understanding of the formation 
and evolution of Pop III stars, the understanding of the sources that reionize the Universe at that epoch. The most 
distant galaxy has been detected at z = 6.96\,\cite{Iye06} while the most distant AGN has been detected at 
$z \sim 6.43$\,\cite{Willott07}. Photometric indications (these galaxies and AGN are too faint to get a spectrum even 
with the very large telescopes) exist of objects with $7 < z <10$; what is really needed is the spectrum in order to have 
not only a certain identification but also the possibility to measure continuum and lines to estimate the population 
and the metal abundance.\\
Swift detected three objects for which the optical follow up evidenced through their spectra very high z objects: 
GRB\,050904 at z = 6.29\,\cite{Kawai06}, GRB\,080913 at z=6.7\,\cite{Greiner09} and GRB\,090423 at 
z = 8.2\,\cite{Salvaterra09,Tanvir09}. The latter hold the record for any celestial object so far observed.\\

The host galaxy of GRB\,050904\,\cite{Berger07} indicate a mass smaller than a few $10^{9}$ solar masses while the 
metal lines\,\cite{Kawai06} call for a rather low metallicity $Z \sim 0.05 Z_{\odot}$. Unfortunately the spectrum 
of GRB\,090423 does not show any detectable emission or absorption line due to the very small signal to noise ratio. 
To make any progress in this field we need to get to the target as soon as possible and obtain reasonably high S/N ratio 
spectroscopic observations.\\

Finally we should clarify the morphology between short and long gamma ray bursts in connection with the physics of 
their formation and the identification of the progenitors. They both have similar characteristics on the decaying light 
curve, naturally referring to those cases in which the light curve of shorts has been observed and can clearly be 
distinguished as two well separated classes only in the Amati relation. Fast follow up will enable (1) collection of 
the first optical afterglow spectrum of a short burst, providing an in-­situ probe of short burst environs and 
possible progenitor signatures; (2) searches for the predicted signature of the decay of radioactive sub--relativistic 
material at early times; and (3) collection of extended afterglow light curves to probe the beaming angle distribution 
of the short bursts via ``jet break'' analyses, a crucial input in estimating merger event rates. So far the spectra of 
these GRBs have been elusive due to their faintness and extremely rapid optical decay. In conclusion to make any further 
significant progress on the GRB physics and related modeling we need a medium size robotic telescope described above.

\section*{Acknowledgments}
This work is supported by ASI grant SWIFT I/011/07/0, by the Ministry of University and Research of Italy 
(PRIN, MIUR, 2007TNYZXL), by MAE and by the University of Milano--Bicocca (Italy).

%

\begin{thebibliography}{0}
\bibitem{Racusin08}
RACUSIN J. ET AL. {\it Nature}, {\bf 455} (2008) 183

\bibitem{Guidorzi10}
GUIDORZI C. ET AL. in preparation

\bibitem{Steele06}
STEELE I.A. ET AL.,  {\it Proc. SPIE}, {\bf 6269}, (2006) 179S

\bibitem{Vitali10}
VITALI F. ET AL., {\it SPIE Astronomical Telescopes and Instrumentation: Observational Frontiers of Astronomy for the New Decade}, Vol.{\bf 7733}
\,(2010)

\bibitem{Campana06}
CAMPANA S. ET AL. {\it Nature}, {\bf 442} (2006) 1008

\bibitem{Mazzali08}
MAZZALI P. ET AL. {\it Science}, {\bf 321} (2008) 1185

\bibitem{Soderberg08}
SODERBERG A. ET AL. {\it Nature}, {\bf 453} (2008) 469

\bibitem{Fryer07}
FRYER C. {\it ApJ}, {\bf 662} (2007) L55

\bibitem{Iye06}
IYE M. ET AL. {\it Nature}, {\bf 443} (2006) 186

\bibitem{Willott07}
WILLOTT C.J. ET AL. {\it AJ}, {\bf 14} (2007) 2435

\bibitem{Kawai06}
KAWAI N. ET AL. {\it Nature}, {\bf 440} (2006) 184

\bibitem{Greiner09}
GREINER J. ET AL. {\it ApJ}, {\bf 693} (2007) 1610

\bibitem{Salvaterra09}
SALVATERRA R. ET AL. {\it Nature}, {\bf 461} (2009) 1258

\bibitem{Tanvir09}
TANVIR N. ET AL. {\it Nature}, {\bf 461} (2009) 1254

\bibitem{Berger07}
BERGER E. ET AL. {\it ApJ}, {\bf 665} (2007) 102

%
%
\end{thebibliography}
\end{document}